\documentclass[prd,twocolumn,showpacs,superscriptaddress,nofootinbib,floatfix,showkeys,10pt]{revtex4-2}

\usepackage{titlesec}
\titlespacing*{\section}{0pt}{3ex plus 1ex minus .2ex}{2ex}
\usepackage{graphicx}
\usepackage{amsmath}
\usepackage{bm}
\usepackage{yhmath}
\usepackage{mathtools}
\usepackage{wasysym}
\usepackage[colorlinks,citecolor=blue,urlcolor=blue,linkcolor=blue]{hyperref}
\usepackage{subfigure}
\usepackage{color}
\usepackage{cases}
\usepackage{subfigure}
\usepackage{times}
\usepackage{dcolumn,booktabs,bm}
\usepackage{slashed}
\usepackage{amsfonts,amssymb,stmaryrd,latexsym,amsmath}
\usepackage{textcomp}
\usepackage{multirow}
\usepackage{cancel}
\usepackage{array}
\usepackage{orcidlink}
\usepackage{enumitem}
\usepackage{booktabs}
\usepackage{tabularx}
\usepackage{caption}

\renewcommand{\arraystretch}{1.8}

\begin{document}


\title{Singly heavy tetraquark resonant states with multiple strange quarks}

\author{Xin-He Zheng\,\orcidlink{0009-0002-2550-331X}}\email{zhengxh@stu.pku.edu.cn}
\affiliation{School of Physics, Peking University, Beijing 100871, China}

\author{Yao Ma\,\orcidlink{0000-0002-5868-1166}}\email{yaoma@pku.edu.cn}
\affiliation{School of Physics and Center of High Energy Physics, Peking University, Beijing 100871, China}

\author{Shi-Lin Zhu\,\orcidlink{0000-0002-4055-6906}}\email{zhusl@pku.edu.cn}
\affiliation{School of Physics and Center of High Energy Physics, Peking University, Beijing 100871, China}

\begin{abstract}

We systematically investigate the S-wave singly heavy tetraquark systems containing two or three strange quarks, $Qs\bar{s}\bar{s}$, $Qn\bar{s}\bar{s}$ and $Qs\bar{s}\bar{n}\left( Q=c,b,n=u,d \right) $, within the constituent quark potential model. We solve the four-body Schrödinger equation using the Gaussian expansion method (GEM) and identify resonances via the complex scaling method (CSM). There are no bound states below the lowest two-meson thresholds. We obtain several compact resonances with $J^P=0^+,2^+$ in $Qs\bar{s}\bar{s}$, and $J^P=2^+$ in $Qn\bar{s}\bar{s}$ and $Qs\bar{s}\bar{n}$. The pole positions are mainly distributed around $7.0–7.2$ GeV (bottom) and $3.7–3.9$ GeV (charm), with widths from a few to several tens of MeV. These resonances decay into $D_s\eta ^\prime ,{D_{(s)}^*}\phi ,{D_s}^*K^*$ and $D_s^*\bar{K}^*$ (and their bottom counterparts), providing targets for future experimental searches.

\end{abstract}

\maketitle

\section{Introduction}~\label{sec:intro}

 Beyond conventional mesons and baryons, quantum chromodynamics (QCD) permits the existence of exotic hadrons such as multiquark states, hybrids, and glueballs ~\cite{Jaffe:1976ig,Jaffe:1975fd,Fritzsch:1973pi}. The investigation of these exotic states provides valuable insights into the nonperturbative dynamics of QCD. Over the past two decades, the observation of numerous exotic-hadron candidates has stimulated extensive theoretical interest, making this field one of the most vibrant in hadronic physics (see Refs.~\cite{Chen:2016qju,Hosaka:2016pey,Esposito:2016noz,Ali:2017jda,Lebed:2016hpi,Guo:2017jvc,Liu:2019zoy,Brambilla:2019esw,Meng:2022ozq,Chen:2022asf,Mai:2022eur} for recent reviews).

Among the exotic hadrons, the tetraquark states attract particular interest. Those with open heavy flavor and strange quarks form a crucial category. A prominent example arises from the anomalous charmed–strange states $D_{s0}^{*}\left( 2317 \right) $~\cite{BaBar:2003oey,CLEO:2003ggt} and $D_{s1}\left( 2460 \right) $~\cite{Belle:2003guh,BaBar:2006eep}, whose discovery challenged the conventional $q\bar{q}$ meson picture and motivated interpretations including compact tetraquarks, hadronic molecules, and mixed configurations~\cite{Chao:2004nb,Liu:2004kd,Zhang:2006hv,Jovanovic:2007bz}. The SELEX's observation of a narrow $D_{sJ}\left( 2632 \right) $~\cite{SELEX:2004drx} further strengthened the tetraquark hypothesis, as its decay patterns are difficult to reconcile within a standard $c\bar{s}$ framework~\cite{Liu:2004kd,Zhang:2006hv}. Early theoretical studies employing coupled-channel formalisms~\cite{Gerasyuta:2008ps,Gerasyuta:2008hs}, diquark–antidiquark model~\cite{Ebert:2010af}, and phenomenological hyperfine interactions~\cite{Jovanovic:2007bz} interpreted these anomalous charmed–strange states as multiquark configurations and suggested that the presence of multiple strange quarks significantly modifies the spectral structure of tetraquarks.

Following the anomalies in the charmed–strange spectrum, the LHCb Collaboration observed two resonances, $T_{cs0}(2900)$ and $T_{cs1}(2900)$, in the $D^-K^+$ spectrum of $B^+ \to D^+D^-K^+$, with $J^P=0^+$ and $1^-$ and quark content $cs\bar u \bar d$~\cite{Aaij:2020hon,Aaij:2020ypa} in 2020. These discoveries, representing the first tetraquark candidates with four distinct flavors, immediately motivated extensive theoretical efforts. Interpretations include near-threshold hadronic molecules~\cite{Molina:2010tx,Chen:2020aos,He:2020jna,Liu:2020orv,Hu:2020mxp,Agaev:2020nrc,Wang:2021lwy,Wang:2023jgb}, compact tetraquark scenarios within various quark models~\cite{Karliner:2020vsi,He:2020jna2,Wang:2020prk,Zhang:2020xtb,Wang:2020xyc,Lu:2020qmp,Tan:2020ldi,Albuquerque:2020ugi,Yang:2021izl,Agaev:2022eea,Liu:2022reg}, and kinematical effects such as triangle singularities or cusps~\cite{Liu:2020orv2,Burns:2020epm}. Their strong decays~\cite{Yu:2017zst,Huang:2020ptc,Xiao:2020ltm}, production mechanisms~\cite{Chen:2020eyu,Lin:2022wmj,Yu:2023}, and possible partner states~\cite{Bayar:2022cwf,Dai:2022htx} have also been investigated, with broader perspectives summarized in recent reviews~\cite{Brambilla:2019esw,Chen:2016qju,Guo:2017jvc,Liu:2019zoy,Meng:2022ozq,Chen:2022asf}. Recently, a quark model study employing the Gaussian expansion and complex scaling methods predicts a resonance around 2906~MeV with a width of about 20~MeV, consistent with the observed $T_{cs0}(2900)$ and supporting its interpretation as a $D^*\bar K^*$ molecular candidate~\cite{Chen:2023syh}.

In addition to the $T_{cs}(2900)$ states, the LHCb Collaboration observed a pair of open-charm tetraquark candidates in the $D_{s}\pi$ channel, namely $T_{c\bar{s}0}(2900)^{++}$ and $T_{c\bar{s}0}(2900)^{0}$, with masses near 2.9~GeV and minimal quark content $c\bar{s}\bar{u}d$ or $c\bar{s}\bar{d}u$~\cite{LHCb:2020bls}. Their masses and discovery channels agree very well with the predictions in Ref. \cite{Chen:2017rhl}. Their measured masses and widths are roughly consistent with the earlier $T_{cs}(2900)$ signals, though the strangeness assignments differ. 
QCD sum rules reproduce the observed mass and width, supporting a compact $[cs][\bar u\bar d]$ interpretation~\cite{Yang:2023evp,Lian:2023cgs}. 
Coupled–channel analyses of $D^*K^*$–$D_s^*\rho$ dynamics predict a pole near threshold and partner states with other quantum numbers~\cite{Duan:2023lcj}. 
Triangle singularities have been proposed as a possible non-resonant explanation of the $T_{c\bar{s}}(2900)$ peak, though only specific loops seem capable of reproducing the observed structure~\cite{Ge:2022dsp}. 
Alternatively, within the framework of effective theory, it is described as a near-threshold $D^*K^*$ molecule, with characteristic decay patterns into $D_s\pi$ and $DK$~\cite{Wang:2023hpp}. 

Together with the subsequent $T_{c\bar{s}}$ signals, the $T_{cs}$ observations have become a key reference for open-flavor exotica, providing a natural bridge to systematic studies of singly heavy tetraquarks within various theoretical approaches. In particular, systems with strange or multi-strange constituents have been explored in relativized quark models and related frameworks, which reveal consistent spectral patterns governed by heavy-quark and SU(3)-flavor symmetries~\cite{Lu:2016zhe,Lu:2020qmp}. In parallel, approaches such as the improved chromomagnetic interaction model~\cite{Guo:2021mja} and algebraic treatments of rotational-vibrational dynamics~\cite{Jalili:2023kmw} provided alternative perspectives on the internal structure of exotic candidates like the $T_{cs}(2900)$. Collectively, these studies suggested that systems with multiple strange quarks exhibit flavor-dependent spectral features compared with their non-strange counterparts.

Despite these achievements, most existing analyses concentrated on ground-state spectra or neglected a rigorous treatment of resonances above thresholds, leaving unresolved questions about the stability and internal structure of such systems.
To address these gaps, we investigate the bound and resonant states of the singly heavy tetraquark systems containing two or three strange quarks ($Qs\bar{s}\bar{s}$, $Qn\bar{s}\bar{s}$, $Qs\bar{s}\bar{n}$) using the AL1 constituent quark potential model. By solving the four-body Schrödinger equation with the Gaussian expansion method~\cite{Hiyama:2003cu} and identifying resonances through the complex scaling method~\cite{Aguilar:1971ve,Balslev:1971vb,aoyama2006complex}, our results provide a unified description of the spectra and structural features of these systems and offer targets for future experimental searches.

This paper is organized as follows. In Sec.~\ref{sec:framework}, we outline the theoretical formalism and computational methods employed. In Sec.~\ref{sec:results}, we present and discuss the numerical results, including mass spectra, structural features, and decay channels of the obtained tetraquark states. Finally, Sec.~\ref{sec:sum} summarizes the main results.

\section{Theoretical framework}~\label{sec:framework}

We employ a constituent quark potential model. The four-body problem is solved with the Gaussian expansion method (GEM), and resonances are identified with the complex scaling method (CSM). This framework has been employed in our earlier studies~\cite{Chen:2023syh,Wu:2024zbx,Yang:2025wqo,Wu:2024euj,Wu:2024hrv,Ma:2024vsi,Wu:2024ocq}.

\subsection{Hamiltonian}~\label{subsec:Hamiltonian}

The nonrelativistic four-body Hamiltonian in the center-of-mass frame is
\begin{align}\label{eq:Hamiltonian}
H=\sum_i^4\left(m_i+\frac{\boldsymbol{p}_i^2}{2m_i}\right)+\sum_{i<j=1}^4 V_{i j}\,,
\end{align}
where $m_i$ and $\boldsymbol{p}_i$
represent the mass and momentum of the $i$-th (anti)quark, respectively. The term $V_{ij}$ describes the two-body interaction between the $i$-th and $j$-th (anti)quark. 

We adopt the AL1 potential, which combines one-gluon exchange and linear confinement,
\begin{align}\label{eq:AL1}
    V_{i j}=-\frac{3}{16}&\lambda_i^c \cdot \lambda_j^c\Big(-\frac{\kappa}{r_{ij}}+\lambda r_{i j}-\Lambda\nonumber\\
    &+\frac{8\pi\kappa'}{3m_{i}m_{j}}\frac{\exp(-r_{ij}^{2}/r_{0}^{2})}{\pi^{3/2}r_{0}^{3}}\boldsymbol{s}_{i}\cdot\boldsymbol{s}_{j}
    \Big),
\end{align}
where $r_0=A\left(\frac{2m_im_j}{m_i+m_j}\right)^{-B}$. $\lambda^c$ represents the SU(3) color Gell-Mann matrix and $\boldsymbol{s}_i$ is the spin operator of quark $i$. The parameters of the potential have been determined by fitting the meson spectra in Ref.~\cite{Silvestre-Brac:1996myf} and are listed in Table~\ref{tab:paraAL1}. The calculated meson masses and root-mean-square (rms) radii in AL1 model are shown in Table~\ref{tab:meson}.

\begin{table}[htbp]
    \centering
    \caption{The parameters in the AL1 quark potential model.}
    \label{tab:paraAL1}
    \begin{tabular*}{\hsize}{@{}@{\extracolsep{\fill}}cccccc@{}}
        \hline\hline
        $ \kappa $ & $ \lambda { [\mathrm{GeV}^{2}]}$ & $ \Lambda {\rm [GeV]} $ & 
        $ \kappa^\prime $ & $ A { [\mathrm{GeV}^{B-1}]}$ & $ B $ \\
        \hline
        0.5069 & 0.1653 & 0.8321 & 1.8609 & 1.6553 & 0.2204 \\
        \hline
        $ m_b \,[\mathrm{GeV}] $ & $ m_c \,[\mathrm{GeV}] $ & 
        $ m_s \,[\mathrm{GeV}] $ & $ m_q \,[\mathrm{GeV}] $ & & \\
        \hline
        5.227 & 1.836 & 0.577 & 0.315 & & \\
        \hline\hline
    \end{tabular*}
\end{table}


\begin{table}[htbp]
	\renewcommand{\arraystretch}{1.4}
	\centering
	\caption{\label{tab:meson} Theoretical masses (MeV) and rms radii (fm) of mesons obtained with the AL1 model, compared with experimental values from Ref.~\cite{ParticleDataGroup:2022pth}.}
	\begin{tabular*}{\hsize}{@{\extracolsep{\fill}}cccc|ccc@{}}
		\hline\hline
		Meson & $m_{\mathrm{Exp.}}$ & $m_{\mathrm{Theo.}}$ & $r^{\mathrm{rms}}_{\mathrm{Theo.}}$ 
		& Meson & $m_{\mathrm{Theo.}}$ & $r^{\mathrm{rms}}_{\mathrm{Theo.}}$ \\
		\hline 
		$K$ & 496 & 491 & 0.59 & $K(2S)$ & 1465 & 1.31 \\
		$K^*$ & 894 & 904 & 0.81 & $K^*(2S)$ & 1643 & 1.42 \\
		$\ensuremath{\eta^{\prime}}$\footnote{For simplicity, we assume no mixing between the $I=0$ $\eta(n \bar{n})$ and the $\eta^{\prime}(s \bar{s})$.} 
		& -- & 714 & 0.54 
		& $\ensuremath{\eta^{\prime}(2S)}$ & 1565 & 1.17 \\ 
		$\ensuremath{\phi}$ & 1020 & 1021 & 0.70 
		& $\ensuremath{\phi(2S)}$ & 1695 & 1.25 \\
		$D$ & 1867 & 1862 & 0.61 & $D(2S)$ & 2643 & 1.23 \\
		$D^*$ & 2009 & 2016 & 0.70 & $D^*(2S)$ & 2715 & 1.27 \\
		$D_s$ & 1968 & 1963 & 0.50 & $D_s(2S)$ & 2661 & 1.02 \\
		$D_{s}^{*}$ & 2112 & 2102 & 0.57 & $D_{s}^{*}(2S)$ & 2720 & 1.06 \\
		$B$ & 5279 & 5294 & 0.63 & $B(2S)$ & 6013 & 1.20 \\
		$B^*$ & 5325 & 5351 & 0.66 & $B^*(2S)$ & 6041 & 1.22 \\
		$B_s$ & 5367 & 5361 & 0.49 & $B_s(2S)$ & 5998 & 0.97 \\
		$B_{s}^{*}$ & 5415 & 5418 & 0.52 & $B_{s}^{*}(2S)$ & 6021 & 0.99 \\
		\hline \hline
	\end{tabular*}
\end{table}

\subsection{Wave function construction}~\label{subsec:wavefunction}

For the S-wave four-quark systems considered here, the total wave function $\psi$ is expressed as an antisymmetric product of its color-spin $\chi$ and spatial $\phi$ components. This is formally represented as:
\begin{equation}\label{eq:Abasis}
\psi=\mathcal{A}\left(\chi \otimes \phi\right),
\end{equation}
where $\mathcal{A}$ is the antisymmetrization operator, ensuring compliance with the Pauli principle for identical fermions.

For the spatial wave function, the Gaussian expansion method (GEM)~\cite{Hiyama:2003cu} is employed. The spatial wave function is expanded in the basis:
\begin{equation}\label{eq:basisSpace}
\phi_{n l m}(\boldsymbol{r})=\sqrt{\frac{2^{l+5 / 2}}{\Gamma\left(l+\frac{3}{2}\right) r_n^3}}\left(\frac{r}{r_n}\right)^l e^{-\frac{r^2}{r_n^2}} Y_{l m}(\hat{r}),
\end{equation}
where the $r_n$ is taken in geometric progression, $r_n=r_0 a^{n-1}$. $Y_{l m}$ is the spherical harmonics.

\begin{figure}[htbp]
  \centering
  \includegraphics[width=0.47\textwidth]{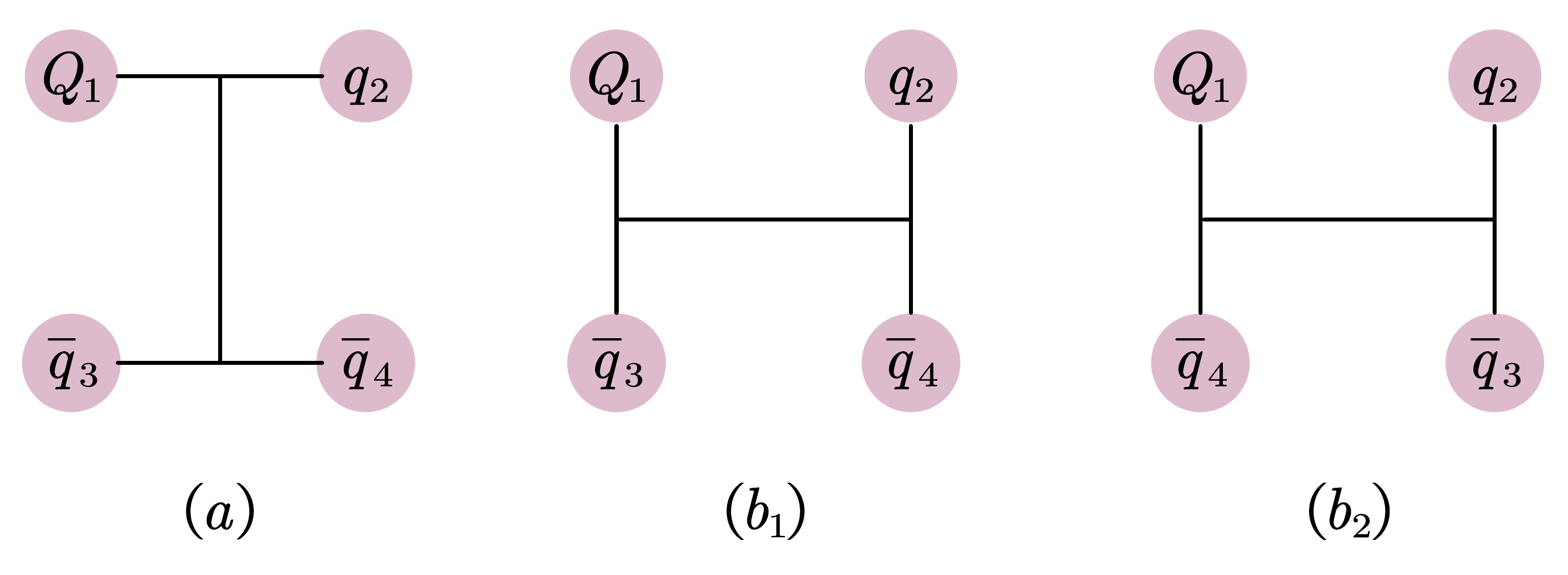} 
  \caption{\label{fig:structure} Two structures of the tetraquark system. (a) diquark-antidiquark structure. (b) dimeson structure. }
    \setlength{\belowdisplayskip}{1pt}
\end{figure}

We describe the internal quark positions of the tetraquark using Jacobi coordinates. In principle, a complete set of bases can be constructed from a single type of Jacobi coordinate. However, in this work we employ only the $S$-wave bases to avoid the complexity of including higher orbital angular momenta. Meanwhile, we include multiple Jacobi coordinate structures to compensate for the omission of higher partial waves. This strategy has been shown effective in previous tetraquark studies~\cite{Meng:2023jqk,Chen:2023syh}. In this work, we include two types of Jacobi coordinates: diquark–antidiquark and dimeson structures, as shown in Fig.~\ref{fig:structure}.

For the flavor wave function, our study focuses on singly heavy tetraquark systems without $n\bar{n}$ pairs.
For the color-spin wave function, we can choose two complete sets
of bases written as
\begin{equation}
\begin{aligned}
&\left\{
\begin{aligned}
\chi _{1_c\otimes 1_c}^{s_1,s_2,S}&=\left[ \left( Q_1\bar{q}_3 \right) _{1_c}^{s_1}\left( q_2\bar{q}_4 \right) _{1_c}^{s_2} \right] _{1_c}^{S},\\
\chi _{8_c\otimes 8_c}^{s_1,s_2,S}&=\left[ \left( Q_1\bar{q}_3 \right) _{8_c}^{s_1}\left( q_2\bar{q}_4 \right) _{8_c}^{s_2} \right] _{1_c}^{S},
\end{aligned}
\right. \\[6pt]
&\left\{
\begin{aligned}
\chi _{\bar{3}_c\otimes 3_c}^{s_1',s_2',S}&=\left[ \left( Q_1q_2 \right) _{\bar{3}_c}^{s_1'}\left( \bar{q}_3\bar{q}_4 \right) _{3_c}^{s_2'} \right] _{1_c}^{S},\\
\chi _{6_c\otimes \bar{6}_c}^{s_1',s_2',S}&=\left[ \left( Q_1q_2 \right) _{6_c}^{s_1'}\left( \bar{q}_3\bar{q}_4 \right) _{\bar{6}_c}^{s_2'} \right] _{1_c}^{S}.
\end{aligned}
\right.
\end{aligned}
\end{equation}
Since both sets are orthogonal and complete, using either set will yield exactly the same results~\cite{Meng:2023jqk}.

\subsection{Complex scaling method}~\label{subsec:method}

Resonant states cannot be handled as ordinary Hermitian eigenvalue problems because their wave functions are not square-integrable. To overcome this difficulty, the complex scaling method (CSM) is often employed. It analytically rotates coordinates and momenta~\cite{Aguilar:1971ve,Balslev:1971vb,aoyama2006complex},
\begin{align}\label{eq:complexRotation}
U(\theta) \boldsymbol{r}=\boldsymbol{r} e^{i \theta}, \quad U(\theta) \boldsymbol{p}=\boldsymbol{p} e^{-i \theta},
\end{align}
yielding a non-Hermitian Hamiltonian
\begin{equation}\label{eq:HamiltonianComplex}
H(\theta)=\sum_{i=1}^4\left(m_i+\frac{p_i^2 e^{-2 i \theta}}{2 m_i}\right)+\sum_{i<j=1}^4 V_{i j}\left(r_{i j} e^{i \theta}\right).
\end{equation}
As a result, the wave functions of resonant states within the angle $2 \theta$ become square-integrable, allowing them to be solved using localized Gaussian bases, similar to bound states.

In the complex energy plane, different types of states are located in characteristic regions:
\begin{itemize}
	\item \textbf{Bound states:} Located along the negative real axis.
	\item \textbf{Continuum states:} Form rays originating from relevant thresholds, with an argument of $\mathrm{Arg}\left( E \right) =-2\theta $.
	\item \textbf{Resonant states:} Identified by complex eigenvalues $E_R = M_R - i\Gamma_R/2$, where $M_R$ is the mass and $\Gamma_R$ is the decay width. Only resonances satisfying $|Arg(E_R)| < 2\theta$ can be resolved under our framework.
\end{itemize}
Both bound and resonant state positions remain invariant with respect to changes in the rotation angle $\theta$~\cite{Lin:2022wmj,Chen:2023eri,Chen:2023syh}.

\subsection{Spatial structure}

The rms radius serves as a probe of spatial configurations, providing a handle to differentiate compact and molecular states. Under CSM, the conventional definition of the rms radius is given by:
\begin{equation}
r_{i j}^{\mathrm{rms,C}} \equiv \operatorname{Re}\left[\sqrt{\frac{\left(\Psi(\theta)\left|r_{i j}^2 e^{2 i \theta}\right| \Psi(\theta)\right)}{\left(\Psi(\theta) \mid \Psi(\theta)\right)}}\right],
\end{equation}
where $\Psi(\theta)$ represents the complex-scaled wave function. The c-product $\left(\phi_n \mid \phi_m\right) \equiv \int \phi_n(\boldsymbol{r}) \phi_m(\boldsymbol{r}) \mathrm{d}^3\boldsymbol{r}$ (omitting the complex conjugate of the "bra" state) ensures the analyticity of the integrand and the stability of the expectation value under varying rotation angles~\cite{Romo:1968tcz}. Although the rms radius from the c-product is generally complex, its real part effectively reflects internal quark clustering of resonant states that are not excessively broad~\cite{homma1997matrix}.

However, for tetraquark systems containing identical quarks, this conventional definition may not reliably distinguish molecular structures due to the antisymmetrization of identical quarks. To address this ambiguity, a non-orthogonal decomposition of the total wave function has been proposed~\cite{Chen:2023syh,Wu:2024euj,Wu:2024hrv}. For a tetraquark, it can be expressed as:
\begin{equation}
\begin{aligned}
	\Psi (\theta )=&\sum_{s_1\ge  s_2}{\left[ \left[ \left( Q_1\bar{q}_3 \right) _{1_c}^{s_1}\left( q_2\bar{q}_4 \right) _{1_c}^{s_2} \right] _{1_c}^{S}\otimes \psi \left( r_1,r_2,r_3,r_4;\theta \right) \right.}\\
	&\left. -\left[ \left( Q_1\bar{q}_4 \right) _{1_c}^{s_1}\left( q_2\bar{q}_3 \right) _{1_c}^{s_2} \right] _{1_c}^{S}\otimes \psi \left( r_1,r_2,r_4,r_3;\theta \right) \right]\\
	=&\mathcal{A} \left[ \sum_{s_1\ge s_2}{\left[ \left( Q_1\bar{q}_3 \right) _{1_c}^{s_1}\left( q_2\bar{q}_4 \right) _{1_c}^{s_2} \right] _{1_c}^{S}}\otimes \psi \left( r_1,r_2,r_3,r_4;\theta \right) \right]\\
	\equiv &\mathcal{A} \Psi _{13,24}(\theta ),\\
\end{aligned}
\end{equation}
where $s_1$ and $s_2$ sum over all spin configurations for the total spin $S$. The decomposed wave function $\Psi _{13,24}$ is then used to define a new rms radius:
\begin{equation}
	r_{ij}^{\mathrm{rms,M}} \equiv \operatorname{Re}\left[\sqrt{\frac{\left(\Psi_{13,24}(\theta)\left|r_{i j}^2 e^{2 i \theta}\right| \Psi_{13,24}(\theta)\right)}{\left(\Psi_{13,24}(\theta) \mid \Psi_{13,24}(\theta)\right)}}\right] .
\end{equation}
This newly defined rms radius provides a more reasonable description of the spatial structure for molecular tetraquark states involving identical quarks. We will adopt this definition in the following sections.

\section{Numerical results}~\label{sec:results}

\subsection{$Qs\bar{s}\bar{s}$}\label{subsec:Qsss}

\begin{figure*}[htbp]
  \centering
  \includegraphics[width=1.0\textwidth]{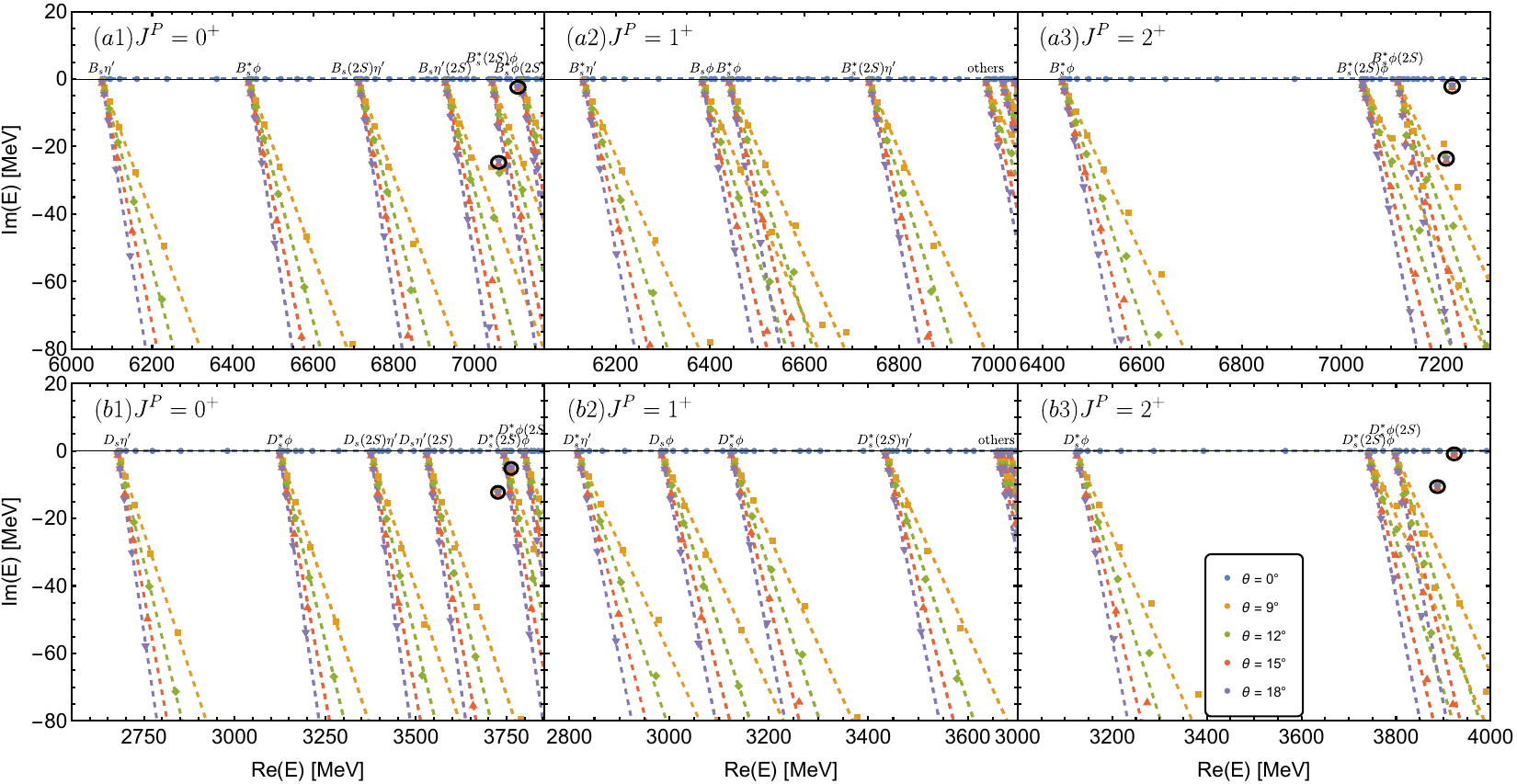} 
  \caption{\label{fig:Qsss}The complex energy eigenvalues of the $Qs \bar{s} \bar{s}$ states in the AL1 potential with varying $\theta$ in the CSM. The dashed lines represent the continuum lines rotating along $\mathrm{Arg}(E)=-2\theta$. The resonant states do not shift as $\theta$ changes and are marked out by the black circles.}
    \setlength{\belowdisplayskip}{1pt}
\end{figure*}

We analyze the S-wave $Qs\bar{s}\bar{s}$ tetraquark states ($J^P=0^+,1^+,2^+$) and present their mass spectra as well as spatial properties. The complex eigenenergies for both $bs\bar{s}\bar{s}$ and $cs\bar{s}\bar{s}$ are obtained using the complex scaling method at several rotation angles ($\theta=9^\circ,12^\circ,15^\circ,18^\circ$). No bound states are found below the lowest thresholds, whereas multiple resonances are identified, as highlighted by black circles in Fig.~\ref{fig:Qsss}. Detailed numerical results, including complex energies, color configuration ratios, and rms radii, are summarized in Table~\ref{tab:Qsss}. For convenience, we denote the resonances as $T_{Qs\bar{s}\bar{s},J^P}\left( M \right) $, where $M$ is the mass.

\begin{table*}[htbp] 
	\centering
	\caption{The complex energies $E=M-i \Gamma / 2$ (in MeV), proportions of different color configurations and rms radii (in fm) of the $b s \bar{s} \bar{s}/c s \bar{s} \bar{s}$ resonant states in the AL1 potential. The last column shows the spatial structure of the states, where C. and M. represent the compact and molecular structure, respectively.}
	\label{tab:Qsss}
	\resizebox{\linewidth}{!}{%
		\renewcommand{\arraystretch}{1.15}
		\small 
		\setlength{\tabcolsep}{4.5pt} 
		\begin{tabular}{@{}l c c c c c c c c c c c c c@{}}
			\hline\hline 
			\multirow{2}{*}{System} & \multirow{2}{*}{$J^{P}$} & \multirow{2}{*}{$M-i \Gamma / 2$} & 
			\multicolumn{4}{c}{Color Configurations} & 
			\multicolumn{6}{c}{Rms Radii} & 
			\multirow{2}{*}{Structure} \\ 
			\cmidrule(lr){4-7} \cmidrule(lr){8-13}
			& & & $\chi_{\overline{3}_c \otimes 3_c}$ & $\chi_{6_c \otimes \overline{6}_c}$ & 
			$\chi_{1_{c}\otimes1_{c}}$ & $\chi_{8_c \otimes 8_c}$ & 
			$r_{Q_1\bar{s}_3}^{\mathrm{rms}}$ & $r_{s_2\bar{s}_4}^{\mathrm{rms}}$ & 		
			$r_{Q_1s_2}^{\mathrm{rms}}$ & $r_{\bar{s}_3\bar{s}_4}^{\mathrm{rms}}$ & 
			$r_{Q_1\bar{s}_4}^{\mathrm{rms}}$ & $r_{s_2\bar{s}_3}^{\mathrm{rms}}$ \\ 
			\midrule 
			
			\multirow{4}{*}{$bs\bar{s}\bar{s}$} & \multirow{2}{*}{$0^{+}$} 
			& $7060-25i$ & 61\% & 39\% & 46\% & 54\% & 0.86 & 0.89 & 0.87 & 0.92 & 0.92 & 1.03 & C. \\ 
			& & $7108-2i$ & 48\% & 52\% & 51\% & 49\% & 0.63 & 1.16 & 0.85 & 1.13 & 1.02 & 1.02 & C. \\			
			\addlinespace[2pt]\cmidrule{2-14}
			& \multirow{2}{*}{$2^{+}$} 
			& $7210-24i$ & 88\% & 12\% & 37\% & 63\% & 0.74 & 1.13 & 0.83 & 0.97 & 0.59 & 1.28 & C. \\ 
			& & $7223-2i$ & 86\% & 14\% & 38\% & 62\% & 0.68 & 1.13 & 0.96 & 0.96 & 0.85 & 1.01 & C. \\
			\midrule
			
			\multirow{4}{*}{$cs\bar{s}\bar{s}$} & \multirow{2}{*}{$0^{+}$} 
			& $3726-12i$ & 55\% & 45\% & 48\% & 52\% & 0.77 & 1.03 & 0.97 & 0.99 & 0.97 & 1.15 & C. \\
			& & $3763-5i$ & 46\% & 54\% & 51\% & 49\% & 0.84 & 0.96 & 0.92 & 1.12 & 1.00 & 1.04 & C. \\
			\addlinespace[2pt]\cmidrule{2-14}
			& \multirow{2}{*}{$2^{+}$} 
			& $3888-11i$ & 79\% & 21\% & 40\% & 60\% & 0.82 & 1.09 & 1.02 & 1.09 & 0.92 & 1.37 & C. \\
			& & $3923-1i$ & 91\% & 9\% & 36\% & 64\% & 0.77 & 1.09 & 1.04 & 0.96 & 0.87 & 1.05 & C. \\
			
			\hline\hline 
		\end{tabular}
	} 
\end{table*}

The mass uncertainties are several tens of MeV, primarily due to intrinsic uncertainties of the constituent quark model. Furthermore, effects from the finite widths of constituent mesons are neglected, and only two–body strong decays are considered. Thus, the theoretical values are expected to underestimate the actual widths.

In tetraquark systems, numerous dimeson thresholds exist, each introducing a branch cut between different Riemann sheets. In practice, the rotation angle must be chosen to ensure the pole of interest is well separated from the nearby continuum eigenvalues. Otherwise, numerical instabilities may occur, as discussed in Refs.~\cite{homma1997matrix,Wu:2024hrv}.

For the S-wave $bs\bar{s}\bar{s}$ system, two resonant states are identified for $J^P=0^+,2^+$. In these cases, no single color configuration dominates. Because no significant differences are observed among the rms radii between each pair of quarks for the four resonant states, all states are classified as compact, in line with our previous numerical experience, which shows that tetraquark systems without valence $u$ or $d$ quarks rarely form bound or near-threshold molecular states and instead favor compact configurations.

For $J^P=0^+$, the $T_{bs\bar{s}\bar{s},0^+}\left( 7060 \right) $ with a width of 50 MeV is very close to the continuum line of the $B_{s}^{*}\left( 2S \right) \phi $ threshold and is therefore numerically less stable compared with $T_{bs\bar{s}\bar{s},0^+}\left( 7108 \right) $ resonance with a width of only 4 MeV.  Both the $\bar{3}_c\otimes 3_c$ and $6_c\otimes \bar{6}_c$ color configurations make significant contributions, indicating that both  interactions within the diquark (or anti-diquark) and between the diquark and anti-diquark pair are important. The former resonance decays into $B_s\eta ^{\prime},B_{s}^{*}\phi ,B_s\left( 2S \right) \eta ^{\prime}$ and $B_s\eta ^{\prime}\left( 2S \right) $, while the latter additionally decays into $B_{s}^{*}\left( 2S \right) \phi $.

For $J^P=1^+$, the dense accumulation of $1S2S$ thresholds above 7.0~GeV prevents a clear separation of possible poles from the continuum states, leading to numerical instabilities. We therefore restrict our analysis to the energy region below this threshold cluster, and within this restricted region no resonance is found.

For $J^P=2^+$, we identify the $T_{bs\bar{s}\bar{s},2^+}\left( 7210 \right) $ with a width of 48 MeV and the $T_{bs\bar{s}\bar{s},2^+}\left( 7223 \right) $ with a width of 4 MeV, decaying into $B_{s}^{*}\phi $ and the corresponding $1S2S$ thresholds. The $2^+$  resonances exhibit significantly higher proportions of $\bar{3}_c\otimes 3_c$ color configuration than of $6_c\otimes \bar{6}_c$ color configuration compared to the $0^+$ resonances. This is obvious from a symmetry perspective: for the $S$-wave $bs\bar{s}\bar{s}$ tetraquark with $J=2$, the $\bar{s}\bar{s}$ pair in an $S$-wave leads to a wave function that is fully symmetric in the combined space–spin–flavor degrees of freedom, while the Pauli exclusion principle restricts the tetraquark to the $\bar{3}_c\otimes 3_c$ color configuration. The reason $\bar{3}_c\otimes 3_c$ doesn't occupy 100$\%$ is that we introduced different Jacobi coordinates. The S-wave bases in the dimeson Jacobi coordinates contribute high partial wave components when projected onto the diquark-antidiquark structure, allowing the color wave function to contain non-antisymmetric components.

Early studies based on relativistic quark model~\cite{Ebert:2010af,Lu:2016zhe}, extended four-body relativized models~\cite{Lu:2020qmp}, improved chromomagnetic interaction models~\cite{Guo:2021mja} and algebraic rotation–vibration approaches~\cite{Jalili:2023kmw} typically assigned the $bs\bar{s}\bar{s}$ spectrum to 6.0–6.9 GeV, but our results push these states beyond 7.0 GeV. A similar feature will also appear in the $cs\bar{s}\bar{s}$, $Qn\bar{s}\bar{s}$ and $Qs\bar{s}\bar{n}$ systems discussed later. The upward shift mainly reflects a methodological difference: earlier quark model studies restricted to localized basis expansions could only treat bound states and thus tended to misidentify continuum excitations as spurious low-lying levels. In contrast, our use of the complex scaling method can distinguish bound, scattering, and resonant states within a unified framework. This enables us to clearly isolate resonance poles above thresholds and avoids the artificial lowering of masses.

For the $cs\bar{s}\bar{s}$ tetraquark, the resonances resemble those of $bs\bar{s}\bar{s}$, differing only in pole positions due to the mass difference between bottom and charm, consistent with heavy-quark symmetry.

For $J^P=0^+$, the $T_{cs\bar{s}\bar{s},0^+}\left( 3726 \right) $ resonance with a width of 24 MeV decays into $D_s\eta ^{\prime},D_{s}^{*}\phi ,D_s\left( 2S \right) \eta ^{\prime}, D_s\eta ^{\prime}\left( 2S \right) $ and is numerically more stable because it is farther away from the continuum line compared with the cousin state $T_{bs\bar{s}\bar{s},0^+}\left( 7060 \right) $ in $bs\bar{s}\bar{s}$. Both $T_{cs\bar{s}\bar{s},0^+}\left( 3726 \right) $ and $T_{cs\bar{s}\bar{s},0^+}\left( 3763 \right) $ exhibit lower $\bar{3}_c\otimes 3_c$ color proportions than their cousins, indicating that the replacement of bottom quark by charm weakens the attractions within the diquark. This observation holds true for most of the states with diquark containing no light quark $u$ or $d$, and it will become even more obvious after replacing the strange quark with charm or bottom~\cite{Wu:2024euj,Wu:2024hrv,Wu:2024zbx}.

For $J^P=1^+$, no resonances are found below the threshold of our interest, as in the case of $bs\bar{s}\bar{s}$. The thresholds above $D_{s}^{*}\left( 2S \right) \eta^\prime $ threshold are very dense. It is very difficult to distinguish the potential resonance candidates buried in the continuum numerically. Furthermore, since the decay widths of the $2S$ or higher excitation mesons are not taken into account, the numerical results of the resonant states above these thresholds are less reliable. Hence, we do not consider these higher resonances.

For $J^P=2^+$, we identify the $T_{cs\bar{s}\bar{s},2^+}\left( 3888 \right) $ resonance with a width of 22 MeV and the $T_{cs\bar{s}\bar{s},2^+}\left( 3923 \right) $ with a width of 2 MeV. The latter resonant state is  dominated by $\bar{3}_c\otimes 3_c$ configuration (91$\%$), which numerically shows that the spatial wave function of diquark is mainly S-wave. Both resonant states couple to $D_{s}^{*}\phi $ and the corresponding $1S2S$ thresholds.

For the $cs\bar{s}\bar{s}$ system, the masses from various quark model investigations in  literature~\cite{Ebert:2010af,Lu:2016zhe,Lu:2020qmp,Guo:2021mja,Jalili:2023kmw,Zhang:2006hv} were between 2.6–3.6 GeV, whereas our calculation only found resonances above 3.7 GeV. 

\subsection{$Qn\bar{s}\bar{s}$}\label{subsec:Qnss}

\begin{figure*}[htbp]
	\centering
	\includegraphics[width=1.0\textwidth]{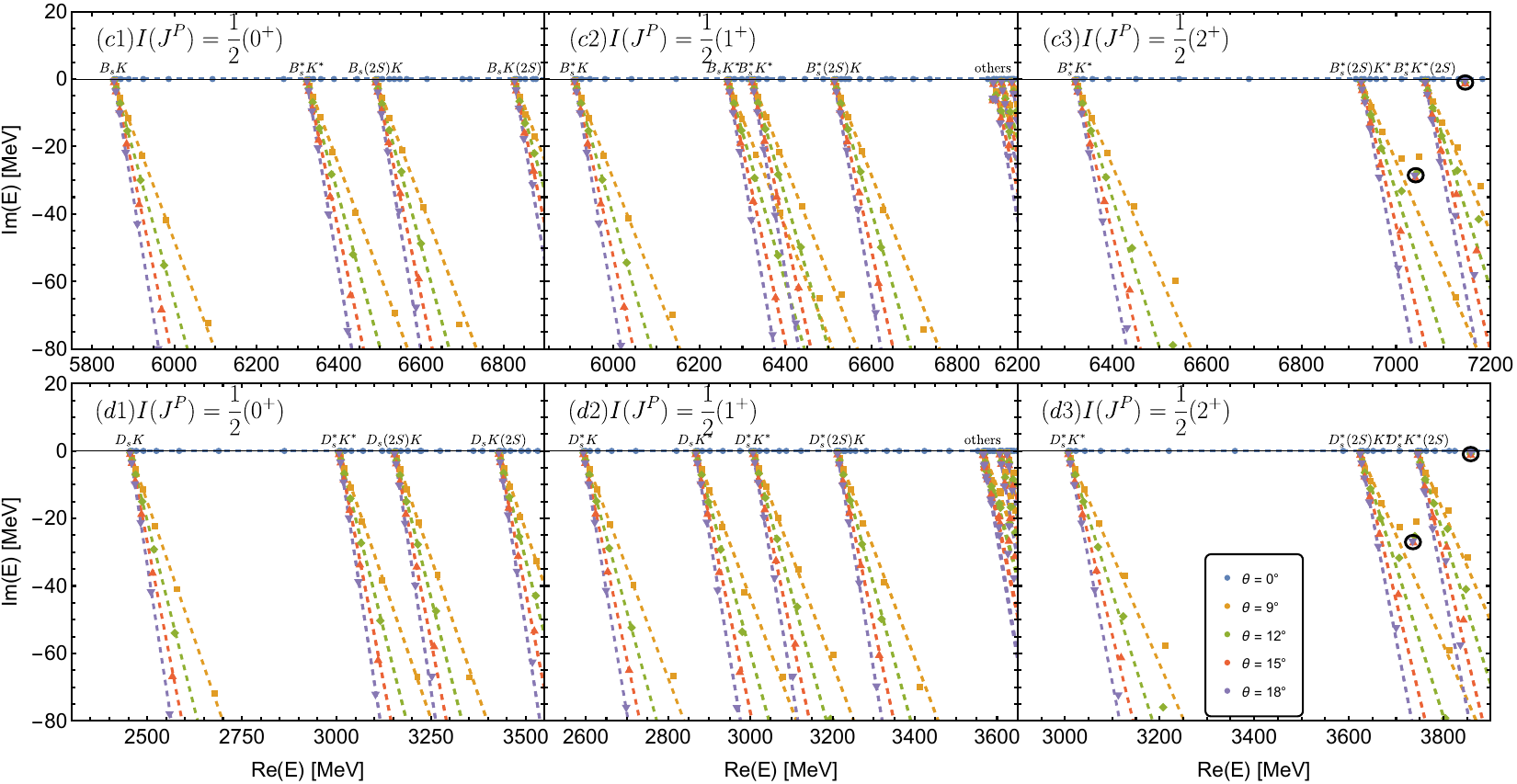} 
	\caption{\label{fig:Qnss}The complex energy eigenvalues of the $Qn \bar{s} \bar{s}$ states in the AL1 potential with varying $\theta$ in the CSM. The dashed lines represent the continuum lines rotating along $\mathrm{Arg}(E)=-2\theta$. The resonant states do not shift as $\theta$ changes and are marked out by the black circles.}
	\setlength{\belowdisplayskip}{1pt}
\end{figure*}

We present the mass spectrum and physical properties of S-wave $Qn\bar{s}\bar{s}$ tetraquarks with $J^{P}=0^{+},1^{+},2^{+}$.
The complex eigenenergies for both $bn\bar{s}\bar{s}$ and $cn\bar{s}\bar{s}$ are shown in  Fig.~\ref{fig:Qnss}. Their complex energies, proportions of different color configurations, and rms radii are summarized in Table~\ref{tab:Qnss}. For convenience, we label the obtained tetraquark resonant states as $T_{Qn\bar{s}\bar{s},J^P}\left( M \right) $.

\begin{table*}[htbp] 
	\centering
	\caption{The complex energies $E=M-i \Gamma / 2$ (in MeV), proportions of different color configurations and rms radii (in fm) of the $b n \bar{s} \bar{s}/c n \bar{s} \bar{s}$ resonant states in the AL1 potential. The last column shows the spatial structure of the states, where C. and M. represent the compact and molecular structure, respectively. The "?" signifies that the rms radii between quarks are numerically unstable as the complex scaling angle $\theta$ changes.}
	\label{tab:Qnss}
	\resizebox{\linewidth}{!}{%
		\renewcommand{\arraystretch}{1.15}
		\small 
		\setlength{\tabcolsep}{4.5pt} 
		\begin{tabular}{@{}l c c c c c c c c c c c c c@{}}
			\hline\hline 
			\multirow{2}{*}{System} & \multirow{2}{*}{$I\left( J^P \right) $} & \multirow{2}{*}{$M-i \Gamma / 2$} & 
			\multicolumn{4}{c}{Color Configurations} & 
			\multicolumn{6}{c}{Rms Radii} & 
			\multirow{2}{*}{Structure} \\ 
			\cmidrule(lr){4-7} \cmidrule(lr){8-13}
			& & & $\chi_{\overline{3}_c \otimes 3_c}$ & $\chi_{6_c \otimes \overline{6}_c}$ & 
			$\chi_{1_{c}\otimes1_{c}}$ & $\chi_{8_c \otimes 8_c}$ & 
			$r_{Q_1\bar{s}_3}^{\mathrm{rms}}$ & $r_{n_2\bar{s}_4}^{\mathrm{rms}}$ & 		
			$r_{Q_1n_2}^{\mathrm{rms}}$ & $r_{\bar{s}_3\bar{s}_4}^{\mathrm{rms}}$ & 
			$r_{Q_1\bar{s}_4}^{\mathrm{rms}}$ & $r_{n_2\bar{s}_3}^{\mathrm{rms}}$ \\ 
			\midrule 
			
			\multirow{2}{*}{$bn\bar{s}\bar{s}$} & \multirow{2}{*}{$\frac{1}{2}\left( 2^+ \right) $} 
			& $7042-29i$ & 21\% & 79\% & 60\% & 40\% & 0.81 & 1.11 & ? & ? & ? & ? & ? \\ 
			& & $7146-1i$ & 86\% & 14\% & 38\% & 62\% & 0.73 & 1.22 & 1.03 & 1.12 & 0.94 & 1.14 & C. \\			
			\addlinespace[2pt]\cmidrule{2-14}
			
			\multirow{2}{*}{$cn\bar{s}\bar{s}$} & \multirow{2}{*}{$\frac{1}{2}\left( 2^+ \right) $} 
			& $3736-27i$ & 14\% & 86\% & 62\% & 38\% & 0.84 & 1.10 & ? & ? & ? & ? & ? \\
			& & $3858-1i$ & 93\% & 7\% & 36\% & 64\% & 0.83 & 1.18 & 1.10 & 1.08 & 0.91 & 1.16 & C. \\
			
			\hline\hline 
		\end{tabular}
	} 
\end{table*}

For the S-wave $bn\bar{s}\bar{s}$ system, two resonant states are identified only for $J^P=2^+$. For $J^P=0^+,1^+$, no resonance states are found in the corresponding range\footnote{One possible pole ($\left( 6882-6i \right) $~MeV) at the upper end of the $1^+$ spectrum is not marked because of the dense thresholds nearby.}.

The lower $T_{bn\bar{s}\bar{s},2^+}\left( 7042 \right) $ resonant state with a width of 58 MeV can decay into $B_{s}^{*}K^*$ or $B_{s}^{*}\left( 2S \right) K^*$. As shown in Fig.~\ref{fig:Qnss} and Table~\ref{tab:Qnss}, this state lies very close to both adjacent continuum spectra, so the rms radii become highly sensitive to the rotation angle and are marked as “?”. In such situations, scattering continua intrude into the spectrum and introduce numerical contamination, which prevents a reliable comparison of the spatial sizes of $T_{bs\bar{s}\bar{s},2^+}(7210)$ and $T_{bn\bar{s}\bar{s},2^+}(7042)$. Nevertheless, their markedly different color configurations make it inappropriate to regard the former as the strange quark partner of the latter. This indicates that the $n \to s$ replacement behaves very differently from the $b \to c$ replacement under heavy quark symmetry. The higher $T_{bn\bar{s}\bar{s},2^+}\left( 7146 \right) $ resonance with a width of only 2 MeV lying above $B_{s}^{*} K^*\left( 2S \right)$ threshold is 86$\%$ dominated by the $\bar{3}_c\otimes 3_c$ configuration. Similar near-zero-width resonances are observed in the following sections, which could be due to the fact that resonances above the $1S2S$ thresholds mix with more sophisticated radial and high partial wave excitations. This results in a weaker overlap with the simple two S-wave meson decay channels in our framework, leading to small widths.

For the S-wave $cn\bar{s}\bar{s}$ system, as in $bn\bar{s}\bar{s}$, only two resonant states are identified for quantum number $J^P=2^+$ and no resonance is found below the threshold of interest for $J^P=0^+,1^+$.

The lower $T_{cn\bar{s}\bar{s},2^+}(3736)$ resonant state with a width of 54~MeV can decay into $D_s^{*}K^{*}$ or $D_s^{*}(2S)K^{*}$. As in the $T_{bn\bar{s}\bar{s},2^+}(7042)$ case, its proximity to adjacent continua causes numerical contamination under CSM, hence some numerically unstable rms radii are marked as "?". The higher $T_{cn\bar{s}\bar{s},2^+}\left( 3858 \right) $ resonance with a width of only 2 MeV lying above $B_{s}^{*} K^*\left( 2S \right)$ threshold is 93$\%$ dominated by $\bar{3}_c\otimes 3_c$ configuration. Given the similarity in pole positions, color-configuration ratios, and rms radii, these two states can be regarded as the charmed partners of $T_{bn\bar{s}\bar{s},2^+}\left( 7042 \right) $ and $T_{bn\bar{s}\bar{s},2^+}\left( 7146 \right) $ respectively.

Conventional approaches, including flavor–spin hyperfine interactions~\cite{Jovanovic:2007bz}, SU(3) multiplet analyses~\cite{Liu:2004kd}, extended relativized quark model~\cite{Lu:2020qmp}, and improved chromomagnetic interaction models~\cite{Guo:2021mja}, suggested states around 2.6–3.5 GeV (charm) and 5.7–6.8 GeV (bottom). In contrast, our spectrum shows no stable poles and instead identifies resonances only at significantly higher energies. The reason has been explained in the discussion of the $bs\bar{s}\bar{s}$ system.

\subsection{$Qs\bar{s}\bar{n}$}\label{subsec:Qssn}

\begin{figure*}[htbp]
	\centering
	\includegraphics[width=1.0\textwidth]{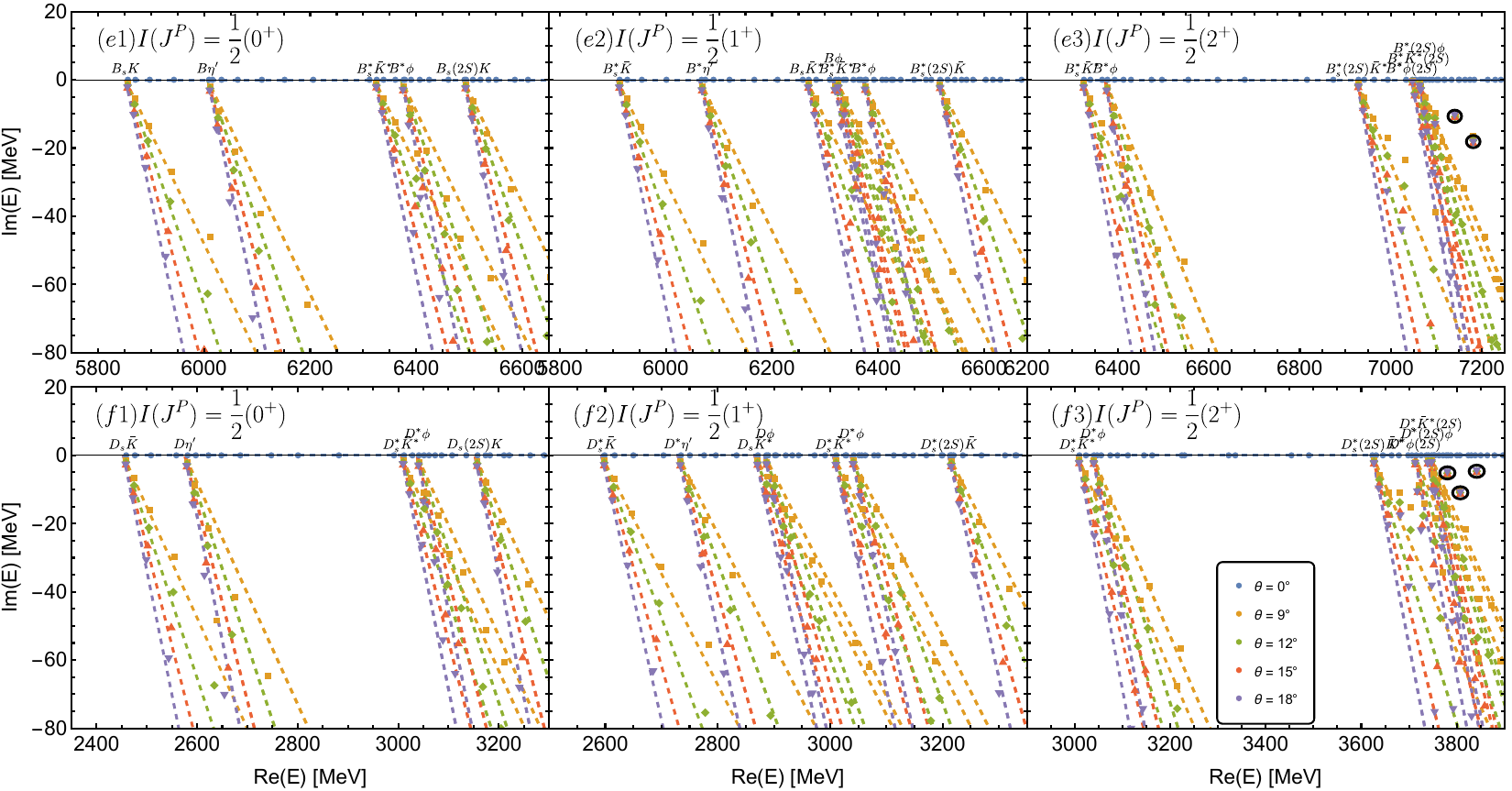} 
	\caption{\label{fig:Qssn}The complex energy eigenvalues of the $Qs \bar{s} \bar{n}$ states in the AL1 potential with varying $\theta$ in the CSM. The dashed lines represent the continuum lines rotating along $\mathrm{Arg}(E)=-2\theta$. The resonant states do not shift as $\theta$ changes and are marked out by the black circles.}
	\setlength{\belowdisplayskip}{1pt}
\end{figure*}

We present the mass spectrum and structural properties of S-wave $Qs\bar{s}\bar{n}$ tetraquarks with $J^{P}=0^{+},1^{+},2^{+}$.
The complex eigenenergies for both $bs\bar{s}\bar{n}$ and $cs\bar{s}\bar{n}$ are shown in  Fig.~\ref{fig:Qssn}. Their complex energies, proportions of different color configurations, and rms radii are summarized in Table~\ref{tab:Qssn}. For convenience, we label the obtained tetraquark resonant states as $T_{Qs\bar{s}\bar{n},J^P}\left( M \right) $.

\begin{table*}[htbp] 
	\centering
	\caption{The complex energies $E=M-i \Gamma / 2$ (in MeV), proportions of different color configurations and rms radii (in fm) of the $b s \bar{s} \bar{n}/c s \bar{s} \bar{n}$ resonant states in the AL1 potential. The last column shows the spatial structure of the states, where C. and M. represent the compact and molecular structure, respectively.}
	\label{tab:Qssn}
	\resizebox{\linewidth}{!}{%
		\renewcommand{\arraystretch}{1.15}
		\small 
		\setlength{\tabcolsep}{4.5pt} 
		\begin{tabular}{@{}l c c c c c c c c c c c c c@{}}
			\hline\hline 
			\multirow{2}{*}{System} & \multirow{2}{*}{$I\left( J^P \right) $} & \multirow{2}{*}{$M-i \Gamma / 2$} & 
			\multicolumn{4}{c}{Color Configurations} & 
			\multicolumn{6}{c}{Rms Radii} & 
			\multirow{2}{*}{Structure} \\ 
			\cmidrule(lr){4-7} \cmidrule(lr){8-13}
			& & & $\chi_{\overline{3}_c \otimes 3_c}$ & $\chi_{6_c \otimes \overline{6}_c}$ & 
			$\chi_{1_{c}\otimes1_{c}}$ & $\chi_{8_c \otimes 8_c}$ & 
			$r_{Q_1\bar{s}_3}^{\mathrm{rms}}$ & $r_{s_2\bar{n}_4}^{\mathrm{rms}}$ & 		
			$r_{Q_1s_2}^{\mathrm{rms}}$ & $r_{\bar{s}_3\bar{n}_4}^{\mathrm{rms}}$ & 
			$r_{Q_1\bar{n}_4}^{\mathrm{rms}}$ & $r_{s_2\bar{s}_3}^{\mathrm{rms}}$ \\ 
			\midrule 
			
			\multirow{2}{*}{$bs\bar{s}\bar{n}$} & \multirow{2}{*}{$\frac{1}{2}\left( 2^+ \right) $} 
			& $7140-11i$ & 92\% & 8\% & 33\% & 67\% & 0.72 & 1.30 & 1.06 & 0.88 & 0.70 & 1.26 & C. \\ 
			& & $7181-18i$ & 48\% & 52\% & 7\% & 93\% & 0.74 & 1.35 & 0.88 & 1.43 & 1.22 & 1.01 & C. \\			
			\addlinespace[2pt]\cmidrule{2-14}
			
			\multirow{3}{*}{$cs\bar{s}\bar{n}$} & \multirow{3}{*}{$\frac{1}{2}\left( 2^+ \right) $} 
			& $3780-5i$ & 59\% & 41\% & 91\% & 9\% & 1.03 & 1.08 & 0.78 & 1.07 & 0.86 & 0.98 & C. \\
			& & $3807-11i$ & 79\% & 21\% & 36\% & 64\% & 0.78 & 1.36 & 0.88 & 0.97 & 0.76 & 1.08 & C. \\
			& & $3842-5i$ & 65\% & 35\% & 15\% & 85\% & 0.84 & 1.31 & 0.93 & 1.28 & 1.14 & 0.93 & C. \\			
			\hline\hline 
		\end{tabular}
	} 
\end{table*}

For the S-wave $bs\bar{s}\bar{n}$ system, two compact resonant states are identified only for $J^P=2^+$. For $J^P=0^+,1^+$, no resonance states are found in the corresponding range.

Both $T_{bs\bar{s}\bar{n},2^+}\left( 7140 \right) $ resonance with a width of 22 MeV and $T_{bs\bar{s}\bar{n},2^+}\left( 7181 \right) $ resonance with a width of 36 MeV possess six decay channels: $B_{s}^{*}\bar{K}^*$, $B^*\phi $, $B_{s}^{*}\left( 2S \right) \bar{K}^*$, $B_{s}^{*}\bar{K}^*\left( 2S \right) $, $B^*\left( 2S \right) \phi $ and $B^*\phi \left( 2S \right) $. The primary color configuration for $T_{bs\bar{s}\bar{n},2^+}\left( 7140 \right) $ is $\bar{3}_c\otimes 3_c$. It is worth noting that the overlap of the three thresholds ($B_{s}^{*}\bar{K}^*\left( 2S \right) $, $B^*\left( 2S \right) \phi $ and $B^*\phi \left( 2S \right) $) makes it impossible to identify the potential resonant states among them.

For the S-wave $cs\bar{s}\bar{n}$ system, three compact resonant states are identified only for $J^P=2^+$. For $J^P=0^+,1^+$, no resonance states are found in the corresponding range.

Although $T_{cs\bar{s}\bar{n},2^+}\left( 3780 \right) $ resonance with a width of 10 MeV has a $c\bar{s}$ rms radius comparable to $D_{s}^{*}\left( 2S \right) $ and the primary color configuration is $1_c\otimes 1_c$, it should not be regarded as a molecular state for the roughly equivalent rms radii between any two of the four quarks.

Previous studies using relativistic coupled-channel formalisms~\cite{Gerasyuta:2008ps,Gerasyuta:2008hs}, relativistic quark model~\cite{Ebert:2010af}, relativized and extended quark model~\cite{Lu:2016zhe,Lu:2020qmp}, improved chromomagnetic interaction models~\cite{Guo:2021mja}, and algebraic rotation–vibration approaches~\cite{Jalili:2023kmw} predicted this type of tetraquark states in the 2.6–3.5 GeV region for charm sector and 5.7–6.8 GeV for bottom, our results reveal no such low-lying structures. Instead, the identified resonances emerge only at higher mass ranges.

\section{Summary and discussion}~\label{sec:sum}

In summary, we have systematically investigated S-wave singly heavy tetraquarks with multiple strange quarks, namely $Qs\bar{s}\bar{s}$, $Qn\bar{s}\bar{s}$, and $Qs\bar{s}\bar{n}$, using the AL1 constituent quark potential model. With the Gaussian expansion method and the complex scaling method, we included both dimeson and diquark–antidiquark structures in a unified framework. We identified poles corresponding to compact resonant states and found no bound states below the lowest two-meson thresholds.

The identified compact resonances in the $Qss\bar{s}$, $Qns\bar{s}$, and $Qs\bar{s}n$ systems appear at $3.7$--$3.9$~GeV in the charm sector and $7.0$--$7.2$~GeV in the bottom sector. These results show how the strangeness and heavy quarks affect the masses and widths of the tetraquark states, and they indicate that the previously proposed low-lying states are not substantiated within our framework of the constituent quark model~\cite{Ebert:2010af,Lu:2016zhe,Lu:2020qmp,Guo:2021mja,Jalili:2023kmw,Zhang:2006hv,Jovanovic:2007bz,Liu:2004kd,Gerasyuta:2008ps,Gerasyuta:2008hs}. The predicted resonances decay into channels such as $D_s\eta ^\prime ,{D_{(s)}^*}\phi ,{D_s}^*K^*,D_s^*\bar{K}^*$, and their bottom counterparts $B_s\eta^\prime ,{B_{(s)}^*}\phi ,{B_s}^*K^*,B_s^*\bar{K}^*$. We therefore encourage targeted searches in $B$ decays and high-energy $pp$ collisions at LHCb or Belle II. And the partial-wave analyses of these final states may reveal the anticipated $0^+$ and $2^+$ quantum numbers of the tetraquark states.




\begin{acknowledgements}

We are grateful to Hui-Min Yang, Yan-Ke Chen, and Wei-Lin Wu for the helpful discussions. This project was supported by the National Natural Science Foundation of China (Grant No. 12475137). The computational resources were supported by High-performance Computing Platform of Peking University.

\end{acknowledgements}

\bibliography{Ref}

\end{document}